# Molecular Motor or Molecular Clock: A Question of Load

Henry Hess

*Department of Biomedical Engineering, Columbia University, USA*

*hhess@columbia.edu*

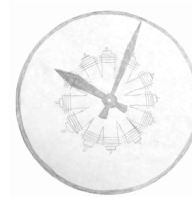

KEYWORDS. Molecular motor, second law of thermodynamics, clock

ABSTRACT. The output of a motor is work, while the output of a clock is information. Here it is discussed how a molecular motor can produce both, work and information, depending on the load. If the ratio of the backward and forward stepping rates of a molecular motor increases exponentially with load, the change in free energy per step can be used to produce only work (at stall force) or only timing information (at zero force), or anything in between.

It could be argued that in the absence of a load, a running motor pointlessly wastes the input energy. However, if the motor runs at a constant speed it can be used as a clock – a machine providing us with information about the passage of time. In the past, clocks were macroscopic entities consuming many $k_BT$ per tick, but circadian clocks[1] and radical clocks[2] demonstrate that timing information can also be generated at the molecular scale. Here, I am interested in the timing information generated by the stepping of a molecular motor, which utilizes an energy input to advance against an external load. An example of such a molecular motor is a kinesin motor protein, which autonomously hydrolyzes ATP and takes 8 nm steps along a microtubule while pulling external loads of up to 7 pN.[3] However, the molecular motor could be driven by other types of energy (e.g. light[4]) and could have an output which is not mechanical work (e.g. charge transport against an electric field) as long as it operates autonomously.[5]

How much timing information does a molecular motor produce? If we disregard the stochastic fluctuations in the duration of each step for the moment, and assume that the motor steps at equal time intervals, a molecular motor step produces the same amount of information as the tick of a watch. For a given observation window (e.g. a day), we can assign events a time stamp defining after which tick it has occurred. If we divide a finite time interval T into N subintervals using a clock stepping N times at a rate $k_+=N/T$, we can create N groups of events and tag each group with a time stamp providing $log_2(N)$ bits of information.

A motor stepping continuously forward at precise time intervals would be a perfect clock, but at the microscopic scale backward steps occur at a rate defined by the free energy change resulting from each step ∆G and the work W performed against the opposing force: $k_-/k_+ = \exp[(\Delta G-W)/k_BT]$. Each backward step will degrade the information in the N time stamps. The first backward step makes the last bit of the time stamp unreliable, and additional backward steps continue this process. If n backward steps occur at a rate of $k_-=n/T$, each time stamp will only provide in average $log_2(N/n) = log_2(k_+/k_-)$ bits.

Of course, the opposing force also reduces the number of forward steps N in a given time interval unless the motor mechanism is a pure power stroke.[6] However, the number of catalytic cycles decreases in proportion to the number of forward steps and the number of significant bits of each time stamp are still given by $log_2(k_+/k_-)$.

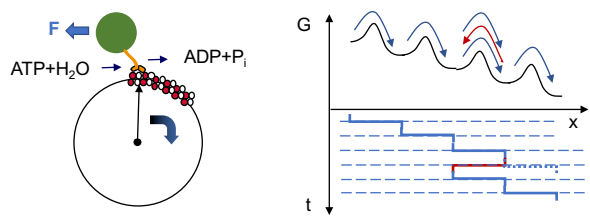

**Figure 1:** A kinesin motor protein undergoes a chemomechanical cycle as it moves along a microtubule against an external load hydrolyzing one ATP molecule for every step. Such a stepping motor provides timing information, which is degraded by backward steps. An opposing force increases the frequency of backward steps.

The ratio of forward and backward rates allows us to relate average information generated per step C= $log_2(N/n)$ to the free energy per step:

$$\Delta G = W + C\, k_B T \ln 2 \qquad (1)$$

This implies that the molecular motor converts the free energy either into timing information or work. Thus even an unloaded motor (F=0 and W=0) delivers something in return for the expenditure of free energy, just as the moving arms of a Swiss watch make the winding worthwhile.

It can be argued that Eq. (1) is merely a restatement of the well-known relationship between free energy, work and forward and backward rates (stated e.g. in [3]):

$$\Delta G = W + k_B T \ln(k_+/k_-) \qquad (2)$$

While the equations are certainly formally equivalent, I would like to emphasize the point of this essay that Eq. (1) describes the operation of a device having aspects of a motor and a clock, while Eq. (2) makes a statement purely about a motor. The physics of a system consisting of an opening and sand under the influence of gravity can be described by mechanics, but it is still worthwhile to point out that this system can be employed as a sand clock.

Looking at a molecular motor as a clock with a load-dependent accuracy is of course closely related to the extensive and insightful recent work of Barato, Pietzonka and Seifert[7-11] which explores the relationship between energy dissipation, efficiencies, and precision of molecular machines. Nevertheless, there are some differences in the present approach. In "Cost and Precision of Brownian Clocks" Barato and Seifert[9] aim to determine the energetic cost to measure a given time interval (e.g. an hour) with a given uncertainty and find that the energetic cost is independent of the unit of clock time (the time per tick), that is a clock measuring an hour with a one minute precision by counting seconds requires just as much energy dissipation as a clock counting minutes. This task however resembles the task of an alarm clock (or a fuse) rather than the task of a wrist watch or wall clock. A wrist watch returns timing information continuously, and adding the capability of measuring seconds with a third arm provides additional information to the wearer, which is of interest here. Secondly, Barato and Seifert account for the randomness in the individual transitions, which additionally degrades the timing information. Here, the perspective of a "long-term" average is taken where the noise introduced by the variability in the timing of each step has vanished. This is of course a serious shortcoming if we time a single event occurring after a small number of steps (alarm clock or fuse). However, for a clock continuously producing timing information over a large number of steps, these fluctuations do average out and a simple connection to the thermodynamic quantities defined for equilibrium states can be made.

Other connections can be made to the work of Jarzynski and colleagues, who illustrated how information can be converted into work by a mechanical demon,[12] and of course to Landauer's seminal insight that information erasure requires a minimum of energy dissipation and the developments resulting from it.[13] It should be noted that information in statistical thermodynamics quantifies uncertainty (with larger uncertainty corresponding to a higher information content), whereas "timing information" as used here reduces uncertainty. Therefore the apparent paradox that both the production of timing information and the erasure of (thermodynamic) information require energy expenditure.

The main message of this short essay is that even an unloaded molecular motor provides a return for an investment in free energy: It can serve as a clock.

ACKNOWLEDGMENT

The author acknowledges financial support from the US Army Research Office under W911NF-13-1-0390 and NSF grant CMMI-1662329.